\documentclass[aps,prl,reprint,amsmath,amssymb,showpacs,superscriptaddress,nofootinbib]{revtex4-2}
\usepackage{CJKutf8}
\usepackage{graphicx}
\usepackage{dcolumn}
\usepackage{bm}
\usepackage{color}
\usepackage{float}
\usepackage{hyperref}

\allowdisplaybreaks[4]

\begin{document}

\title{Abnormal Bifurcation of the Double Binding Energy Differences and Proton-Neutron Pairing: Nuclei Close to $N=Z$ Line from Ni to Rb}

\author{Y. P. Wang}
\affiliation{State Key Laboratory of Nuclear Physics and Technology, School of Physics, Peking University, Beijing 100871, China}
\author{Y. K. Wang}
\affiliation{State Key Laboratory of Nuclear Physics and Technology, School of Physics, Peking University, Beijing 100871, China}
\author{F. F. Xu}
\affiliation{State Key Laboratory of Nuclear Physics and Technology, School of Physics, Peking University, Beijing 100871, China}
\author{P. W. Zhao}
\email{pwzhao@pku.edu.cn}
\affiliation{State Key Laboratory of Nuclear Physics and Technology, School of Physics, Peking University, Beijing 100871, China}
\author{J. Meng}
\email{mengj@pku.edu.cn}
\affiliation{State Key Laboratory of Nuclear Physics and Technology, School of Physics, Peking University, Beijing 100871, China}
\affiliation{Center for Theoretical Physics, China Institute of Atomic Energy, Beijing, 102413, China}

\date{\today}

\begin{abstract} 

The recently observed abnormal bifurcation of the double binding energy differences $\delta V_{pn}$ between the odd-odd and even-even nuclei along the $N=Z$ line from Ni to Rb has challenged the nuclear theories. 
To solve this problem, a shell-model-like approach based on the relativistic density functional theory is established, by treating simultaneously the neutron-neutron, proton-neutron, and proton-proton pairing correlations both microscopically and self-consistently. 
Without any \textit{ad hoc} parameters, the calculated results well reproduce the observations, and the mechanism for this abnormal bifurcation is found to be due to the enhanced proton-neutron pairing correlations in the odd-odd $N=Z$ nuclei, compared with the even-even ones. 
The present results provide an excellent interpretation for the abnormal $\delta V_{pn}$ bifurcation, and provide a clear signal for the existence of the proton-neutron pairing correlations for nuclei close to the $N=Z$ line.

\end{abstract}

\maketitle

\date{today}

The double binding energy difference $\delta V_{pn}$, is an important mass filter for atomic nuclei, and has been frequently used to isolate the residual proton-neutron (\textit{pn}) interaction \cite{ZhangJY1989PLB,Brenner1990PLB}. 
It is closely related to many nuclear structure phenomena, such as the onset of collectivity and deformation \cite{Talmi1962RMP,Federman1977PLB,Casten1985PRL,Cakirli2006PRL}, the evolution of the underlying shell structure \cite{Heyde1985PLB}, and the phase transition behavior in nuclei \cite{Federman1977PLB,Federman1978PLB}.
The study of $\delta V_{pn}$, particularly along the line of the nuclei with equal numbers of protons and neutrons, is of great importance to deepen our understanding of the nuclear force \cite{Warner2006Nature}.
For example, the considerable enhancement of $\delta V_{pn}$ for $N=Z$ nuclei in the light $sd$ shell has been regarded as the fingerprint for the Wigner's SU(4) symmetry of the nuclear force \cite{VanIsacker1995PRL}. 
For heavier nuclei in the $sd$ shell and even lower $fp$ shell, the decrease of $\delta V_{pn}$ with mass number is related to the breaking of the Wigner's SU(4) symmetry due to the increasing spin-orbit and Coulomb interactions.

For the upper $fp$-shell nuclei, however, the $\delta V_{pn}$ shows quite puzzling behavior. On one hand, restrengthening $\delta V_{pn}$ values have been observed \cite{Schury2007PRC,Mardor2021PRC} for the odd-odd nuclei, and this phenomenon might be attributed to the restoration of the pseudo-SU(4) symmetry \cite{VanIsacker1999PRL}, the enhanced overlaps of the proton and neutron wave functions \cite{Brenner2006PRC,ChenL2009PRL,Cakirli2010PRC} or the nuclear deformation \cite{Cakirli2006PRL,Bonatsos2013PRC}. On the other hand, a recent experiment has reported an abnormal bifurcation, namely, opposite evolving trends of $\delta V_{pn}$ with mass number for the even-even and odd-odd $N=Z$ nuclei from Ni to Rb \cite{WangM2023PRL}, which cannot be understood by the aforementioned physical mechanisms and, thus, brought severe challenges to the theoretical models, including the macroscopic-microscopic models \cite{Duflo1995PRC,Moller1995ADNDT,Kawano2006NST,WangN2014PLB,Moller2016ADNDT}, the shell model \cite{Honma2005EPJA}, and the density functional theories (DFTs) \cite{GengLS2005PTP,Kortelainen2012PRC,Goriely2013PRC}.
Note that the macroscopic-microscopic models and DFTs are quite successful for a global description of nuclear masses over the whole nuclear chart.
Their failure in describing the observed $\delta V_{pn}$ bifurcation indicates that important physics may be missing in the current nuclear models.

The valence-space in-medium similarity renormalization group calculations imply that the three-nucleon force has a significant impact on the behavior of $\delta V_{pn}$, but the obtained amplitudes of the $\delta V_{pn}$ bifurcation between the even-even and odd-odd $N=Z$ nuclei are dramatically overestimated \cite{WangM2023PRL}. 
The inclusion of a phenomenological Wigner term in the macroscopic-microscopic models \cite{Duflo1995PRC,Moller1995ADNDT,Kawano2006NST,WangN2014PLB,Moller2016ADNDT} and DFT \cite{Goriely2013PRC} also results in a $\delta V_{pn}$ bifurcation between the even-even and odd-odd $N=Z$ nuclei, but the $\delta V_{pn}$ for the odd-odd nuclei are systematically underestimated \cite{WangM2023PRL}. 

For $N=Z$ nuclei, it is very important to take into account the \textit{pn} pairing correlations. 
In particular, for odd-odd $N=Z$ nuclei, the last valence neutron and proton could be paired due to the \textit{pn} pairing correlations, which are responsible for the phenomenological Wigner terms \cite{Satula1997PLB2,Satula1997PLB,Bentley2014PRC,Negrea2014PRC}.
To solve the puzzle of the $\delta V_{pn}$ bifurcation, a microscopic model which could treat the neutron-neutron (\textit{nn}), proton-proton (\textit{pp}) and \textit{pn} pairing correlations simultaneously and self-consistently is necessary. The blocking effects for odd-odd nuclei should be treated carefully.

The nuclear DFT starts from a universal density functional and has achieved great successes in describing many nuclear phenomena \cite{Vautherin1972PRC,Decharge1980PRC,Ring1996PPNP,Vretenar2005PR,Niksic2011PPNP,MengJ2016Relativistic}. 
It is a promising framework to consider the \textit{pn} pairing correlations in a microscopic way. 
In the conventional Bardeen-Cooper-Schrieffer and Bogoliubov methods, the particle number conservation is violated and the Pauli blocking effects in odd-nucleon systems cannot be treated exactly \cite{Ring2004The}. 
The shell-model-like approach (SLAP), also known as particle number conserving method, treats the pairing correlations and the blocking effects exactly by diagonalizing the many-body Hamiltonian in a properly truncated many-particle configuration (MPC) space with good particle number \cite{ZengJY1983NPA,Volya2001PLB}.
It has been implemented in the relativistic \cite{MengJ2006FPC,ChenWeiChia2014PRC,LiuL2015PRC,ShiZ2018PRC,WangYP2023PLB,FFXu2023IJMPE} and nonrelativistic \cite{Pillet2002NPA,LiangWY2015PRC} DFTs to treat the \textit{nn} and \textit{pp} pairing correlations and widely used to investigate both the nuclear ground-state and excited-state properties.
Nevertheless, a self-consistent treatment of the \textit{pn} pairing correlations is missing.

In this Letter, based on the relativistic DFT (RDFT), a SLAP is developed, which allows a microscopic and self-consistent treatment of the \textit{nn}, \textit{pp} and \textit{pn} pairing correlations simultaneously. 
The developed approach will be applied to investigate the abnormal $\delta V_{pn}$ bifurcation for the $N=Z$ nuclei from Ni to Rb. 

In the SLAP based on RDFT (RDFT-SLAP), the many-body Hamiltonian reads
\begin{equation}
\label{many-body-Hamiltonian}
\hat{H}=\hat{H}_0+\hat{H}_{\text{pair}},
\end{equation}
where $\hat{H}_0$ is the one-body part, and $\hat{H}_{\text{pair}}$ is the pairing part. The one-body part reads, 
\begin{equation}
\hat{H}_0=\sum_{k>0}\left[\varepsilon_k^{\pi}\left(a_k^{\dagger}a_k+a_{\bar{k}}^{\dagger}a_{\bar{k}}\right)+\varepsilon_k^{\nu}\left(b_k^{\dagger}b_k+b_{\bar{k}}^{\dagger}b_{\bar{k}}\right)\right],
\end{equation}
where $\bar{k}$ represents the time conjugate of the state $k$, and $\varepsilon_k^{\pi(\nu)}$ are the single-proton (neutron) energies obtained from the Dirac equation,
\begin{equation}
\label{Dirac-equation}
\left[-i\boldsymbol{\alpha}\cdot\boldsymbol{\nabla}+\beta(m+S)+V\right]\psi_k=\epsilon_k\psi_k.
\end{equation}
Here, the scalar field $S$ and vector field $V$ are connected in a self-consistent way to the scalar and vector densities, for details see Ref. \cite{MengJ2006FPC}. The pairing part reads, 
\begin{equation}
\hat{H}_{\text{pair}}=\sum_{T_z=0,\pm 1}\hat{H}_{\text{pair}}^{T_z},\qquad \hat{H}_{\text{pair}}^{T_z}=-G\sum_{k,k^{\prime}>0}^{k\neq k^{\prime}}P_{k,T_z}^{\dagger}P_{k^{\prime},T_z},
\end{equation}
where $G$ is the effective pairing strength, and $k\neq k^{\prime}$ means that the self-scattering for the nucleon pairs is forbidden \cite{MengJ2006FPC}. The \textit{nn} and \textit{pp} pair creation operators are $P_{k,1}^{\dagger}=b_{k}^{\dagger}b_{\bar{k}}^{\dagger}$ and $P_{k,-1}^{\dagger}=a_{k}^{\dagger}a_{\bar{k}}^{\dagger}$ for $T_z=\pm 1$, and the \textit{pn} pair creation operator is $P_{k,0}^{\dagger}=\frac{1}{\sqrt{2}}\left(b_{k}^{\dagger}a_{\bar{k}}^{\dagger}+a_{k}^{\dagger}b_{\bar{k}}^{\dagger}\right)$ for $T_z=0$.

The nuclear wave functions are expressed as
\begin{equation}\label{wavefunction}
\left|\Psi\right\rangle=\sum_{i,\{s_k\}} C_i^{\{s_k\}}\left|\text{MPC}_i^{\{s_k\}}\right\rangle.
\end{equation}
The many-particle configurations $|\text{MPC}_i^{\{s_k\}}\rangle$ with exact proton number $Z$ and neutron number $N$ are expressed as $\left|l_1 l_2\cdots l_N m_1 m_2 \cdots m_Z\right\rangle=b_{l_1}^{\dagger}b_{l_2}^{\dagger}\cdots b_{l_N}^{\dagger}a_{m_1}^{\dagger}a_{m_2}^{\dagger}\cdots a_{m_Z}^{\dagger}|0\rangle$, and the corresponding configuration energy is denoted as $E_{i}$. Here $s_k$ represents the eigenvalue of the seniority operator $\hat{s}_k$ for the state $k$ \cite{Richardson1966PR}, and it is a good quantum number. The expansion coefficients $C_i^{\{s_k\}}$ and, thus, the wave functions are determined by diagonalizing the Hamiltonian $\hat{H}$ in the MPC space. 

Note that the obtained wave function $\left|\Psi\right\rangle$ is used to determine the occupation probabilities for the single-particle states, and thus the nucleon densities should be updated, which in turn determines the scalar and vector fields $S$ and $V$ in the Dirac equation \eqref{Dirac-equation}. Therefore, the full framework should be solved iteratively to achieve the self-consistency \cite{MengJ2006FPC,ShiZ2018PRC,WangYP2023PLB}. Once a self-consistent solution is obtained, one can calculate the pairing energy, 
\begin{equation}
\begin{split}
E_{\text{pair}}&=\langle \Psi|\hat{H}_{\text{pair}}|\Psi\rangle\\
&=\sum_{ij}C_i^*C_j\langle\text{MPC}_i|\hat{H}_{\text{pair}}|\text{MPC}_j\rangle, \\
\end{split}
\end{equation}
which is added to the total energy of RDFT. 

In this work, the relativistic density functional PC-PK1 \cite{ZhaoPW2010PRC} is adopted.
The Dirac equation \eqref{Dirac-equation} is solved in the three-dimensional harmonic oscillator basis in Cartesian coordinates \cite{Niksic2014CPC} with ten major shells, and the quadrupole deformation including triaxiality is considered self-consistently.

\begin{figure*}[ht]
  \centering
  \includegraphics[width=0.8\linewidth]{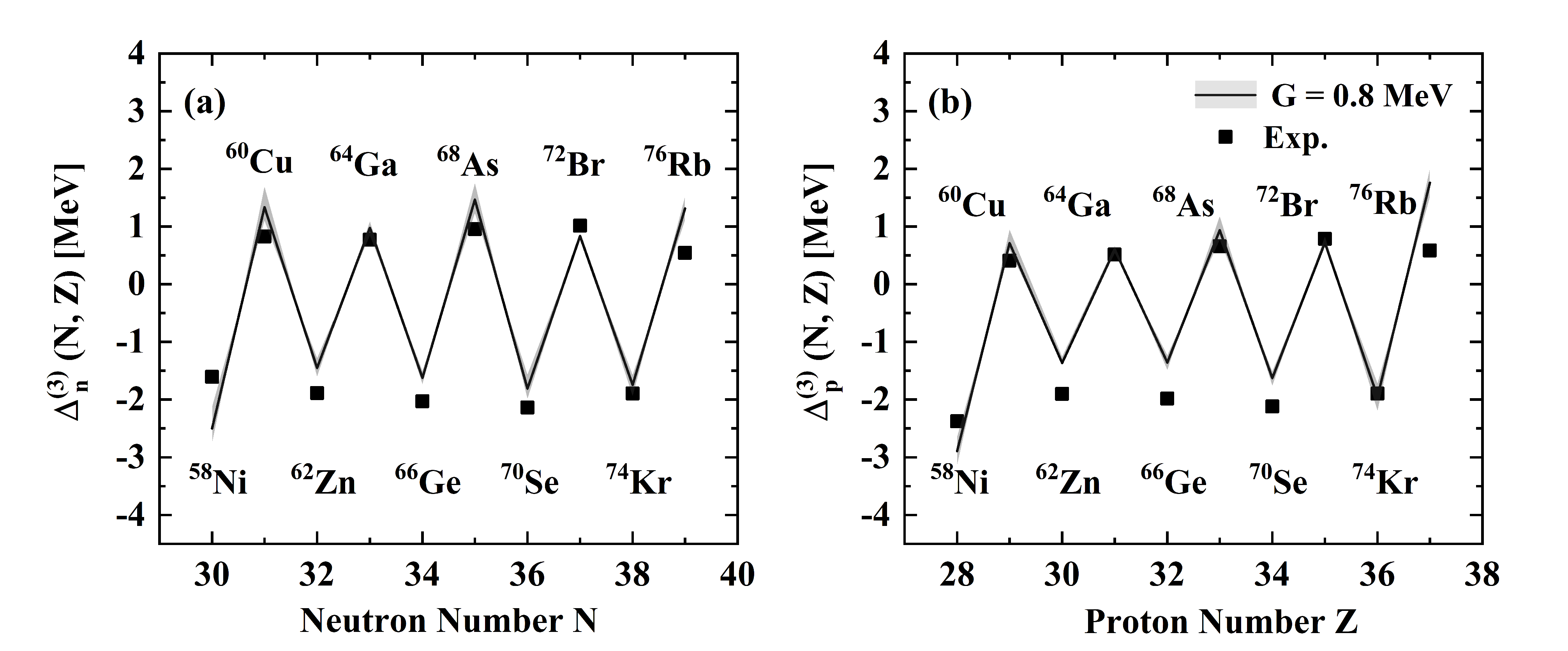}
  \caption{Odd-even mass differences $\Delta_\text{n}^{(3)}$ (a) and $\Delta_\text{p}^{(3)}$ (b) for the $N=Z+2$ nuclei from Ni to Rb calculated by RDFT-SLAP (lines) in comparison with the     
           data (symbols). The gray bands correspond to the odd-even mass differences for the pairing strength $G$ varying by $10\%$.}
  \label{fig1} 
\end{figure*}

The dimension of the MPC space and the corresponding pairing strength $G$ are determined by the odd-even mass differences of the $N=Z+2$ nuclei from Ni to Rb. In Fig. \ref{fig1}, the calculated odd-even mass differences $\Delta_\text{n}^{(3)}(N,Z)$$=$$\left[B(N-1,Z)+B(N+1,Z)\right]/2$$-B(N,Z)$ and $\Delta_\text{p}^{(3)}(N,Z)$$=$$\left[B(N,Z-1)+B(N,Z+1)\right]/2$$-B(N,Z)$ \cite{Bender2000EPJA} are shown, in comparison with the experimental ones extracted from AME’20 \cite{WangM2021CPC,Kondev2021CPC}.
The experimental odd-even mass differences are well reproduced by the calculation with the pairing strength $G=0.8$ MeV. The corresponding MPC space is truncated by $E_{\text{cut}} = 16$ MeV, which means only the MPCs with the energies $E_i \leq 16$ MeV are included in the model space. In addition, a variation of the pairing strength by $10\%$ does not change the odd-even mass differences $\Delta_{\text{n}}^{(3)}$ and $\Delta_{\text{p}}^{(3)}$ significantly. 

\begin{figure*}[ht]
  \centering
  \includegraphics[width=0.8\linewidth]{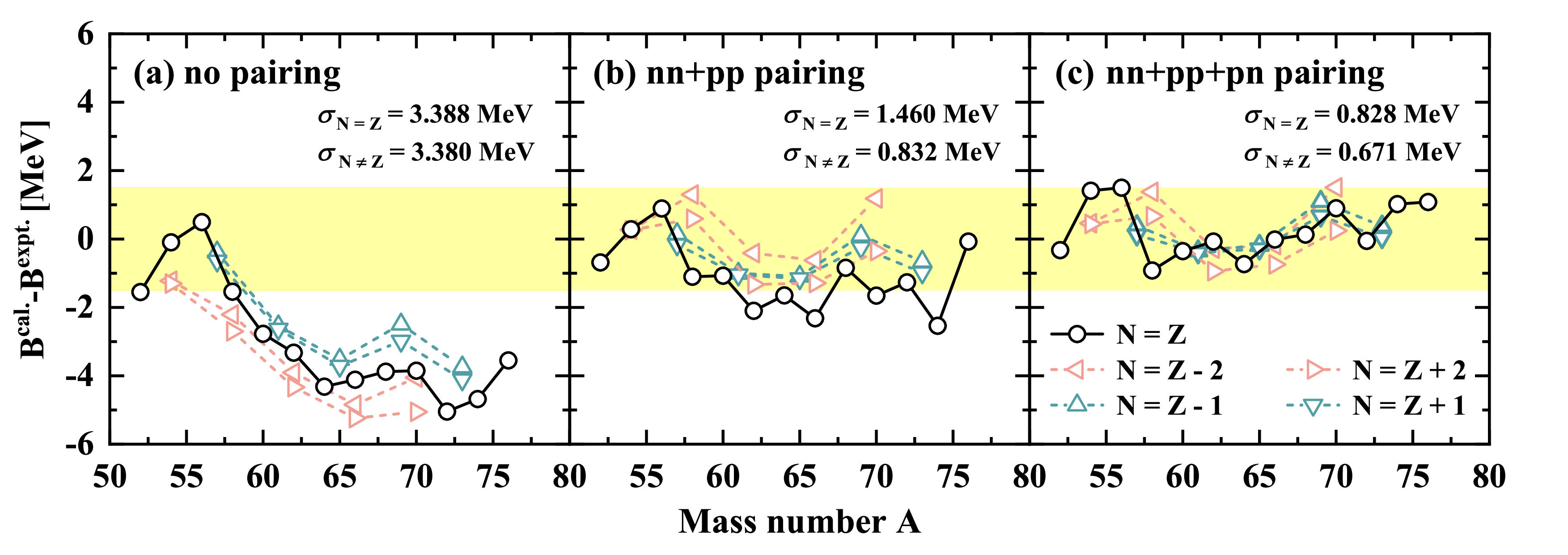}
  \caption{ Differences between the calculated binding energies and the data for the $N=Z, Z\pm 1, Z\pm 2$ nuclei around Ni and Rb region without pairing (a), with the \textit{nn} and \textit{pp} pairing (b), and with the \textit{nn}, \textit{pp} and \textit{pn} pairing (c). The root-mean-square deviation for the $N=Z$ ($N\neq Z$) nuclei $\sigma_{N=Z}$ ($\sigma_{N\neq Z}$) is also given.}
  \label{fig2}
\end{figure*}

With the pairing strength $G$ thus determined, the binding energies for the $N=Z, Z\pm 1, Z\pm 2$ nuclei around Ni and Rb region are calculated. 
The differences between the calculated binding energies and the data \cite{WangM2021CPC,Kondev2021CPC} are shown in Fig. \ref{fig2}.
As shown in Fig. \ref{fig2} (a), without the pairing correlations, the deviations between the calculated binding energies and the data are large. The root-mean-square (rms) deviations are 3.388 MeV for $N=Z$ nuclei and 3.380 MeV for $N\neq Z$ ones. 
After the inclusion of the \textit{nn} and \textit{pp} pairing, as shown in Fig. \ref{fig2} (b), the descriptions of the binding energies are improved. The rms deviation for $N=Z$ nuclei changes to 1.460 MeV, and for $N\neq Z$ nuclei changes to 0.832 MeV. 
After including the \textit{pn} pairing, as shown in Fig. \ref{fig2} (c), the agreements become better. The rms deviation is 0.832 MeV for $N=Z$ nuclei and 0.671 MeV for $N\neq Z$ nuclei \cite{Cu58}.
These results illustrate the importance of the pairing correlations for nuclei near the $N=Z$ line, in particular the \textit{pn} pairing correlations. 

\begin{figure*}[htbp!]
  \centering
  \includegraphics[width=0.4\linewidth]{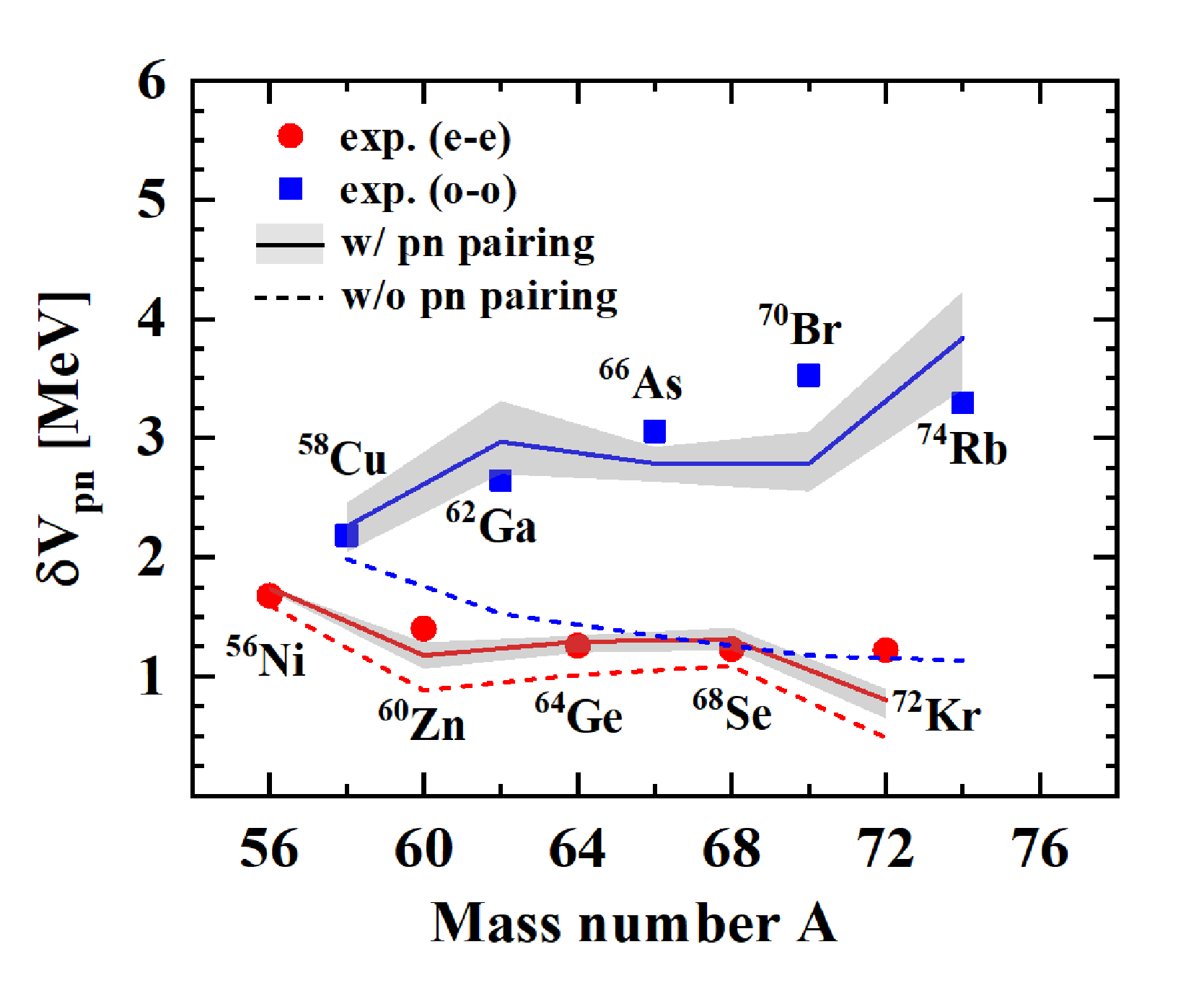}
  \caption{Calculated $\delta V_{pn}$ for the even-even (red lines) and odd-odd (blue lines) $N=Z$ nuclei from Ni to Rb with (solid
           lines) and without (dashed lines) the \textit{pn} pairing, in comparison with the data (symbols) \cite{WangM2023PRL}. The gray bands correspond to the results with the pairing strength $G$ varying by 10\%.}
  \label{fig3}
\end{figure*}

From the binding energies, the $\delta V_{pn}$ can be extracted as \cite{VanIsacker1995PRL}  
\begin{equation}
\label{Vpn-ee}
\begin{split}
\delta V_{{pn}}^{\text{ee}}(N,Z)=\frac{1}{4}&\left[B(N,Z)-B(N-2,Z)-B(N,Z-2)\right.\\
                                 &\left.+B(N-2,Z-2)\right],\\
\end{split}
\end{equation}
for even-even nuclei with $N=Z$, and
\begin{equation}
\label{Vpn-oo}
\begin{split}
\delta V_{{pn}}^{\text{oo}}(N,Z)=&\left[B(N,Z)-B(N-1,Z)-B(N,Z-1)\right.\\
                                 &\left.+B(N-1,Z-1)\right],\\
\end{split}
\end{equation}
for odd-odd nuclei with $N=Z$.
The extracted theoretical results and data \cite{WangM2023PRL} are shown in Fig. \ref{fig3}.
The experimental $\delta V_{pn}^{\text{ee}}$ (red circles) decrease smoothly with the mass number, while the $\delta V_{pn}^{\text{oo}}$ (blue squares) exhibit a distinct tendency; this is the challenging puzzle of the abnormal $\delta V_{pn}$ bifurcation observed in Ref. \cite{WangM2023PRL}.
When the \textit{nn}, \textit{pp} and \textit{pn} pairing correlations are taken into account simultaneously (solid lines), the calculated results well reproduce the evolution of $\delta V_{pn}$ for both the odd-odd and even-even nuclei.
The agreement remains even by changing the pairing strength by 10\%.
In contrast, if the \textit{pn} pairing correlations are switched off (dashed lines), the calculated results cannot reproduce the bifurcation.
As shown clearly in Fig. \ref{fig3}, the successful reproducing for the abnormal $\delta V_{pn}$ bifurcation is due to the enhancement of the $\delta V_{pn}$ for the odd-odd $N=Z$ nuclei by the \textit{pn} pairing correlations.

\begin{figure*}[htbp!]
  \centering
  \includegraphics[width=0.8\linewidth]{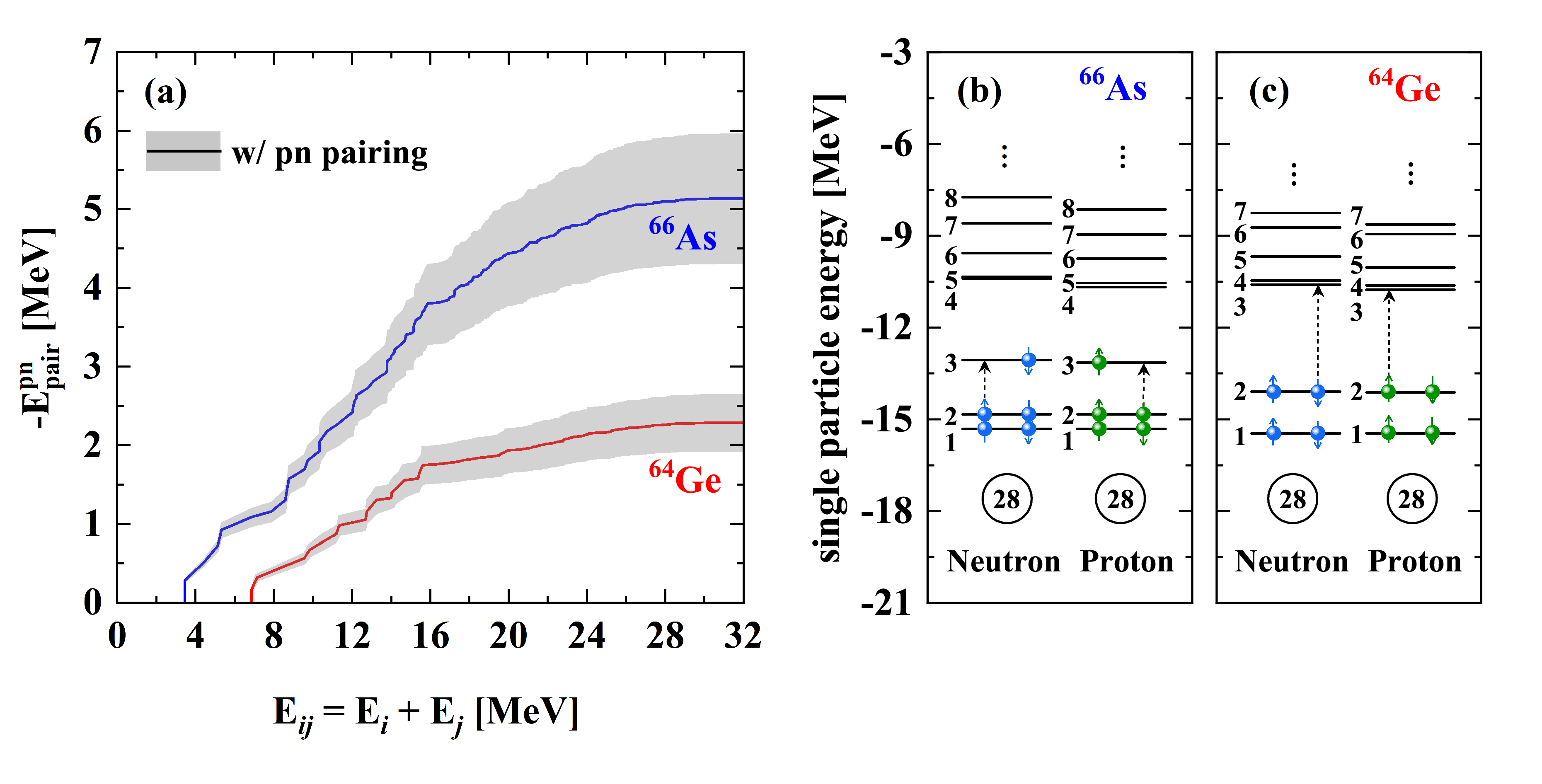}
  \caption{(a) Calculated \textit{pn} pairing energies $E_{\text{pair}}^{\text{pn}}$ as functions of the sums of the configuration energies for the $i$th and $j$th MPC, $E_i+E_j$, for $^{66}$As and $^{64}$Ge. The gray bands correspond to the results for the pairing strength $G$ varying by 10\%. (b) Single-particle energies for the odd-odd nucleus $^{66}$As. The single-proton levels are renormalized to the first single-neutron level above the $N=Z=28$ shell. The lowest-energy MPC and the lowest excitation with nonvanishing contribution to the \textit{pn} pairing energy $E_{\text{pair}}^{\text{pn}}$ are schematically shown. (c) Same as (b), but for the even-even nucleus $^{64}$Ge.}
  \label{fig4}
\end{figure*}

To further understand why the \textit{pn} pairing correlations have more significant influence on the $\delta V_{pn}$ for the odd-odd nuclei as compared to the even-even ones, in Fig. \ref{fig4} (a), the calculated \textit{pn} pairing energies, $E_{\text{pair}}^{\text{pn}}$ $=$ $\langle \Psi|\hat{H}_{\text{pair}}^{T_z=0}|\Psi\rangle$ $=\sum_{ij}C_i^*C_j\langle\text{MPC}_i|\hat{H}_{\text{pair}}^{T_z=0}|\text{MPC}_j\rangle$, are shown as functions of the sums of the configuration energies for $N=Z$ odd-odd nucleus $^{66}$As and even-even nucleus $^{64}$Ge.
For $^{66}$As, the nonvanishing value of $E_{\text{pair}}^{\text{pn}}$ starts at about 3.5 MeV, while for $^{64}$Ge, about 6.9 MeV. 
This can be understood from the lowest MPC and the lowest excitation contributing to the \textit{pn} pairing energy for $^{66}$As and $^{64}$Ge as shown in Figs. \ref{fig4} (b) and \ref{fig4} (c).
With the increase of ${E}_{i}+E_j$, the \textit{pn} pairing energy for $^{66}$As is significantly larger than that for $^{64}$Ge.
Changing the pairing strength $G$ by 10\% will influence the \textit{pn} pairing energy by around 1 MeV for $^{66}$As, and by around 0.4 MeV for $^{64}$Ge.
These results in Fig. \ref{fig4} suggest that the \textit{pn} pairing correlations have more influence on the odd-odd nuclei than on the even-even nuclei, and this explains why the \textit{pn} pairing correlations would significantly enhance the $\delta V_{pn}$ for the odd-odd $N=Z$ nuclei in Fig. \ref{fig3}, and thus result in the abnormal $\delta V_{pn}$ bifurcation.

In summary, a shell-model-like approach is developed based on the relativistic density functional theory, which allows a microscopic and self-consistent treatment of the neutron-neutron, proton-proton and proton-neutron pairing correlations simultaneously.
The challenging puzzle of the abnormal $\delta V_{pn}$ bifurcation between the odd-odd and even-even nuclei along the $N=Z$ line from Ni to Rb is found to be originated from the proton-neutron pairing correlations.
The proton-neutron pairing correlations would significantly enhance the $\delta V_{pn}$ for odd-odd $N=Z$ nuclei, and thus result in the $\delta V_{pn}$ bifurcation.
The proton-neutron pairing correlations improve the description of the masses not only for $N=Z$ nuclei, but also for these nuclei near the $N=Z$ line.
These conclusions remain true even if the pairing strength is changed by 10\%.
The present results provide an excellent interpretation to the challenging puzzle of the abnormal $\delta V_{pn}$ bifurcation, and provide a clear signal for the existence of the proton-neutron pairing correlations for $N=Z$ nuclei.

\begin{acknowledgments}

This work was partly supported by the National Natural Science Foundation of China (Grants No. 11935003, No. 12105004, No. 12141501), the High-performance Computing Platform of Peking University, and the State Key Laboratory of Nuclear Physics and Technology, Peking University.

\end{acknowledgments}


\begin{thebibliography}{10}
\bibitem{Cu58}
The expectation values for the isospin operator $\hat{T}$ \cite{Ring2004The} are calculated for the ground states of all these nuclei, and they agree with the experimental data, except for $^{58}$Cu. For $^{58}$Cu,
in consistent with the data, the state with $\langle\Psi|\hat{T}|\Psi\rangle=0$ is chosen as the ground state.
\end{thebibliography}
\end{document}